# Distribution of doped Mn at the ∑3 (112) grain boundary in Ge

W. Wang,[1,2] S. L. Zhang,[1] Z. X. Tian,[2] and W. Xiao[2,*]

[1]*Department of Applied Physics, Xi'an Jiaotong University, Xi'an 710049, China*

[2]*School of Materials Science & Engineering, University of Science and Technology Beijing, Beijing 100083, China*

Using first-principles density functional theory method, we have investigated the distribution and magnetism of doped Mn atoms in the vicinity of the $\Sigma 3$ (112) grain boundary in Ge. We find that at low concentration, the substitutional sites are energetically favorable over the interstitial ones for Mn. The binding energy of Mn varies with lattice sites in the boundary region, and hence a non-uniform distribution of Mn nears the boundary. However, the average of their segregation energy is quite small, thus no remarkable grain boundary segregation of Mn is predicted. Due to volume expansion at the grain boundary, the spin polarization of Mn is slightly enhanced. Overall, we find that the magnetism of Mn-doped Ge is not sensitively dependent on the grain structure.

**I. INTRODUCTION**

Diluted magnetic semiconductors (DMS) have been investigated extensively due to their potential application in spintronic devices.[1,2,3,4] Doping semiconductor materials such as GaAs[5] and Ge[6] with transition metal atoms can give rise to ferromagnetic coupling between these atoms. In addition, magnetic ions at the cation sites might provide efficient hole doping which helps to modify the concentration of charge carriers. Although more activity is focused on group III-Mn-V materials, there is steadily increasing interest in Ge-based DMS, driven by their potential integration in standard Si semiconductor technology.[7,8,9,10,11,12] In order to achieve highly magnetic properties of efficient spin injection into DMS, increasing the effective dilution of the transition-metal magnetic dopants is necessary. The low-temperature molecular beam epitaxy, a non-equilibrium growth technique is often used to grow DMS materials, by which avoiding Mn precipitation and hence extending Mn concentration limit. However, the price paid for this is the introduction of a large number of defects.[13] More unintended defects form at higher Mn doping because of the tendency of the material toward self-compensation, even under nonequilibrium growth conditions.

It is known that near the free surfaces, the doped Co atoms show different behavior from in the rutile bulk,[14] and grain boundaries (GB) modify the electronic properties by introducing interfacial electronic states and Schottky barriers. A work recently done by Geming *et al.* demonstrated that in Co doped rutile $TiO_2$ GB can enhance the interaction by providing convenient bond angles for ferromagnetic superexchange.[15]

The carrier mediated ferromagnetism in transition-metal-doped IV and III-V semiconductors[16] is quite different from the non-carrier-mediated ferromagnetism in metal oxides such as $TiO_2$[17,18,19] and $SnO_2$.[20] Then one question arises: How would GBs in transition metal doped IV and III-V semiconductors influence the electronic and magnetic properties of the material? The Σ3 (112) GB is one of the dominant boundaries commonly observed in IV and III-V semiconductor films. It has two types: one has coincidence-site-lattice (CSL) without rigid-body translation (RBT), mainly observed in diamond,[21,22] the other has RBT, often found in Ge. Papon and Petit proposed a (2×1) model characterized by reconstructed bonds along the [110], which double the periodicity along this direction.[23] The RBT is $0.091a_0$ along $[\bar{1}1\bar{1}]$ and $0.045a_0$ in $[\bar{1}12]$ ($a_0$ is the cubic lattice constant).

To evaluate the GB stability in Ge and possible segregation of doped Mn atoms, as well as its influence on magnetic properties, we have conducted first-principles density functional theory calculations on the Σ3 (112) GB of Mn doped Ge. The calculated total binding energies of the GB system with Mn located in different sites at low concentration suggest that Mn distribution in the vicinity of the GB is not uniform. Nevertheless, since the average segregation energy near the GB is negligible, we may expect no noticeable Mn segregation in this region. Interestingly, we find that the average magnetization of Mn is strengthened in the GB region, as a result of the volume expansion and distortion of the lattice. The remainder of this paper is organized as follows. In Sec. II, the details of the computational method are described.

In Sec. III, we present the calculations of the grain boundary formation energy with discussion of its stability. In the presence of Mn, we have studied the segregation behavior and spin polarization. Finally, we give a short summary in Sec. IV.

## II. METHODLOGY

We have conducted the first-principles density functional theory calculations using Vienna Ab initio Simulation Package (VASP).[24,25] We used projector augmented wave (PAW) method to describe the electron-ion interaction.[26] The exchange correlation between electrons was treated with generalized gradient approximation (GGA) in the Perdew-Burke-Ernzerhof (PBE) form.[27] We used a cutoff energy of 270 eV for the plane-wave basis. The conjugate-gradient method was employed for the electronic minimization.[28,29] The experimental lattice constant of pure Ge crystal, 5.66 Å, is adopted, as was done by Zhao et al. in a previous PBE calculation.[30] The geometry optimization for each slab used to model the GB was continued until the forces on all the atoms were converged to less than 0.02 eV Å$^{-1}$.

The atomic structure of $\Sigma$ 3 (112) GB in Ge is known from atomic-resolution high-voltage transmission electron microscopy observation.[31,32] We show in Fig. 1 how to build up such a GB with RBT.

**FIG. 1.**

To set up a supercell to model this GB, We need two grains (slabs) that are thick enough to render a bulklike environment in their inner part. Besides, on the side opposite to the grain boundary, we should leave enough space for vacuum. Then, how thick should the grains be? We display in Fig. 2 the calculated increase in total binding energy when adding one more atomic layer to a Ge (112) slab. It is apparent that a slab consisting 12 or more layers can reduce surface-surface interaction down to 0.02eV. To avoid dealing with any possible surface reconstruction, we fix all the atoms in the slab at the corresponding bulk positions.

**FIG. 2.**

The supercell used to simulate the $\sum 3$ (112) GB in Ge in the present work is illustrated in Fig. 3. It contains 112 atoms (28 layers) and has double periodicity along [110] direction. A (3×4×1) Monkhorst-Pack Brillouin-zone sampling is adopted. In order to deal conveniently with the periodic boundary condition, an array of vacuum regions of 10 Å each are inserted in between neighboring slabs to decouple the neighboring slabs. We fixed the atom in four outermost Ge layers at the corresponding bulk positions.

**FIG. 3.**

Note that there are two possible RBTs for a ∑3 (112) GB in Ge. One has the same periodicity along the [110] direction as the perfect crystal, whereas the other has the double periodicity along the same direction. The reconstruction in the [110] direction can eliminate the dangling bond in the RBT (2×1) model. Therefore it is expected to be more stable than the RBT (1×1) model. Using high resolution electron microscope technique, Bourret and Bacmann observed the ∑3 (112) GB in Ge and revealed the GB reconstructed along the [110] direction.[33, 34]

## III. RESULTS AND DISCUSSION

In order to understand the reconstruction of the Σ3 (112) GB in Ge along [110], we have performed electronic structure calculations on this GB for both the (2×1) and (1×1) structures. The unit cell used to model the RBT (2×1) structure consists of 28 layers with four Ge atoms per layer, and the RBT (1×1) structure with two Ge atoms per layer. A (3×4×1) and (3×8×1) Monkhorst-Pack Brillouin-zone sampling are adopted for RBT (2×1) and the RBT (1×1) respectively. Our calculations show that the total energy of a supercell modeling the RBT (2×1) structure is 0.01 eV/Å$^2$ lower than that of the supercell for the (1×1) model.

**FIG. 4.**

In order to examine the change in chemical bonding upon reconstruction along [110] at atom 4, we plot in Fig. 4a and Fig. 4b the total valence electron density on the (112) plane in the RBT (1×1) and RBT (2×1). The contours start from 0.02 e/a.u.$^3$ and increase successively by a factor of $\sqrt{2}$. It is seen clearly that one chemical bond between atom 4 appears when it goes from RBT (1×1) to RBT (2×1) through reconstruction. As a consequence, the total energy of the system is lowered and hence an enhancement of stability. From Fig. 4b, it is apparent that the strength of reconstructed bond 4-4 is weaker than that of the Ge bulk bond 4-9. Numerically, the 4-4 bond length is 5.3% longer than that of the Ge-Ge in bulk. Our calculations show that the deviations (from perfect crystal) in Ge-Ge bond-length near the GB in the RBT (2×1) model are from -2.22% to +4.86%, and the bond-angle deviations range from -13.68° to +18.13° measured from the tetrahedral angle, an indication of a mild distortion.

Having determined the atomic structure of the pure GB, we now discuss the energetics of Mn atoms near the GB at low concentration limit. In reference to the B site, we list in Table I the calculated total energy of a supercell depicted in Fig. 3 with one Ge replaced by one Mn atom. Such an energy difference is the so called segregation energy, $E_{\text{Seg}}$. A negative $E_{\text{Seg}}$ means GB segregation is energetically favorable. Each calculation supercell is denoted by the site where Mn is located.

**TABLE I**

From Table I we find that the Mn atoms at 1, 3, 4, 5, 7, and 9 sites have lower energy, while the systems with Mn atoms at 2, 6, 8, 10, 11, 12, 13, 14, 16, and I sites have higher energy than in the bulk. Overall, our calculations indicate that at low concentration, the distribution of Mn is not uniform in the vicinity of the GB. However, no noticeable segregation will occur on average. As for the in interstitial site at the GB core, the binding energy of Mn with the matrix is 1.0 eV lower than that in the bulk.

The RBT model containing periodically repeated GB without vacuum will not be affordable in the present study. Therefore, we constructed the RBT supercell model by containing just one boundary and a vacuum buffer. In such a model, the two grains are thick enough to render a bulklike environment in their inner part. We note that the surface states may influence the magnetism of the Mn. The electronic defect states typically exist in a covalent material with free surfaces which contain dangling bonds. As a consequence, these electronic defect states will show up near the Fermi energy. There may be spurious interaction with Mn states if they are within the same energy windows of the density of states. Hence, a countercheck of the stability and the magnetism with a H-terminated model for random positions is necessary. We randomly chose the 3, 4, 6 and 8 systems to compare the segregation energy and magnetic moment using the H-terminated RBT model with a vacuum layer. The calculated results are included in Table II. We find the corresponding values of stability and magnetic moments for these propotypical positions using the H-terminated RBT model are very close to our previous calculations. Therefore, we can conclude that the surface states caused by dangling bonds have only marginal influence on the stability and magnetism of Mn on the Ge grain boundary.

**TABLE II**

In the study of Co segregation on Σ5 (113) [-110] GB in anatase $TiO_2$, Gemming *et al.*[15] found that the elastic deformation plays an important role in the segregation. Here we want to see if this is also the case in Mn-doped Ge. Again, we chose the 3, 4, 6 and 8 systems, to compare the orderings of total energy and elastic deformation energy. The elastic deformation energy $E_{el}$ is defined as the energy release during geometry optimization when Mn is replaced back by Ge, which is the scale of distortion accompanied by Mn substitution for Ge. In this way, GB specific local deviations from the bulk elastic properties can be rationalized. The $E_{el}$ can be obtained from the following steps: First, we relaxed geometries of the Mn-substituted supercells, and then carried out fixed-geometry calculations without Mn doping. The $E_{el}$ is the difference between the total energy of such a distorted supercell and the total energy of the fully relaxed pure supercell. The results of $E_{el}$ are listed in Table III.

**TABLE III**

Finally, we will discuss the magnetism of Mn near the GB. In Table I, we have reported that the local magnetic moment of Mn ($M_{Mn}$) in the atomic sphere of Mn (radius 1.30 Å) and the average Mn-Ge bond length ($L_{Mn-Ge}$) at the Σ3 (112) GB in Ge, where the $L_{Mn-Ge}$ denotes the average bondlength. On the whole, we observe an enhanced local magnetic moment associated with enlarged bondlength. It suggests that the variation of the Mn-Ge bondlength is responsible for the change in local magnetic moment of Mn. As we know, the crystal field surrounding magnetic impurities directly affects the magnetism and the strength of crystal field can be determined by the ligand number, the symmetry of complex, and the bond lengths between the ligands and the central cation. In order to gain insights into the Mn-Ge bonding, we employed one primitive unit cell containing two atoms with one Ge atom replaced by Mn (Mn content: 50%).[35] Such a high Mn concentration cannot be easily reproduced experimentally due to the low solubility of Mn impurities in semiconductors, but we believe a study of local magnetic moment of Mn as a function of the Mn-Ge bond length in that system could shed some light on understanding the interaction between Mn and Ge. Here the bond length can be tuned by changing the corresponding lattice constant. We have calculated the local magnetic moment of Mn versus the bond length in a primitive cell (see Fig. 5). As expected, the results demonstrate clearly that the magnetic moment of Mn increases as the bond length increases and will converge toward to $5\mu_B$/Mn. This could be understood from the fact that the hybridization between Mn $3d$ orbitals and N $2p$ orbital decreases as the bond length increases, and finally, the local magnetic moment of Mn will converge to $5\mu_B$/Mn which is the maximum of an isolated Mn atom. Combining with the calculated results displayed in the Table I, one can readily understand that there is a wide variety of magnetic moment of Mn in the GB regions. The minimum magnetic moment value 3.10 $\mu_B$ (Mn at site 6 ) corresponds to the smallest bond-length 2.42 Å, and the maximum magnetic moment value 3.98 $\mu_B$ (Mn at interstitial site ) associated with the largest bond-length 2.87 Å. The magnetic moments in other cases fall in between 3.10 $\mu_B$ and 3.98 $\mu_B$. It should be pointed out that the coincidence of the changes in bondlength and moments is not exact, since a mild distortion, i.e. John-Teller distortion, in tetrahedral crystal field in the grain boundary regions would affect the split of 3$d$ degenerate states whose effect on the magnetism cannot be ignored.

**Fig. 5**

From Tables I and III, we can find that similar to the case of Co doped anatase $TiO_2$, substitutions introducing less distortion corresponds to more stable configuration. This means that the lattice distortion is a decisive factor for Mn segregation on the Σ3 (112) GB in Mn doped Ge. The magnetization of Mn on the Σ3 (112) GB is different from that in the bulk environment. In the latter case, Mn carries a local magnetic moment of 3.29 $\mu_B$; whereas near the GB, the spin magnetic moment inside the atomic sphere of Mn with a radius of 1.3Å ranges from 3.10 $\mu_B$ to 3.98$\mu_B$.

## IV. SUMMARY

We have performed a systematic study on both the atomic structure of a clean Σ3 (112) grain boundary in Ge and the distribution and spin polarization of low-concentration Mn near this boundary, using first-principles density functional theory calculations. We confirmed the previous experimental observation that due to the occurrence of reconstructed bonds, the RBT (2×1) grain boundary structure is more stable than the RBT (1×1) structure. At low concentration, simulated by one Mn in a 112-atom cell, the binding energy of Mn varies with lattice sites in the boundary region. Therefore, the distribution of Mn nears the boundary is not uniform. Nevertheless, the average segregation energy for the lattice sites within eight (112) layers away from the boundary is quite small. As a result, there will be no remarkable grain boundary segregation of Mn. Remarkably, an enhancement of average magnetic moment is found for Mn in the boundary regions, partly due to the fact that Mn-Ge bond lengths at the grain boundary are longer than those in the bulk. On the other hand, the distortion of tetrahedral crystal field in the GB regions also contributes to the variation of the magnetic moment. Since GB regions make only a small portion of the total volume, that the magnetism of Mn-doped Ge will not be sensitively dependent on the grain structure of a polycrystalline material.

**ACKOWLEDGMENTS**

We are grateful to W. T. Geng for initiating this work. We thank J. L. Nie for valuable discussion. This work was supported by the Keygrant Project of Chinese Ministry of Education (No. 708082), NSFC (Grant No. 50831002), NCET (Grant No. 06-0080), RFDP and SRF-ROCS-SEM. The Calculations were performed on the Quantum Materials Simulation of USTB.

TABLE I. The segregation energy, $E_{Seg}$ (eV/cell), spin magnetic moment in an atomic sphere of Mn with a radius of 1.3Å, $M_{Mn}$, and the average Mn-Ge bond length $L_{Mn-Ge}$ (Å). The systems are denoted by the atomic sites where Mn atoms are located (See Fig. 3 for illustration).

| System | 1 | 2 | 3 | 4 | 5 | 6 |
|---|---|---|---|---|---|---|
| $E_{Seg}$ | -0.04 | 0.02 | -0.14 | -0.30 | -0.04 | 0.22 |
| $M_{Mn}(\mu_B)$ | 3.30 | 3.45 | 3.65 | 3.52 | 3.34 | 3.10 |
| $L_{Mn-Ge}$ (Å) | 2.47 | 2.51 | 2.47 | 2.49 | 2.49 | 2.42 |
| System | 7 | 8 | 9 | 10 | 11 | 12 |
| $E_{Seg}$ | -0.05 | 0.02 | -0.06 | 0.08 | 0.04 | 0.10 |
| $M_{Mn}(\mu_B)$ | 3.31 | 3.32 | 3.30 | 3.24 | 3.30 | 3.23 |
| $L_{Mn-Ge}$ (Å) | 2.45 | 2.45 | 2.43 | 2.43 | 2.43 | 2.41 |
| System | 13 | 14 | 15 | 16 | I | B |
| $E_{Seg}$ | 0.23 | 0.17 | 0.00 | 0.18 | 1.00 | 0.00 |
| $M_{Mn}(\mu_B)$ | 3.34 | 3.25 | 3.26 | 3.35 | 3.98 | 3.29 |
| $L_{Mn-Ge}$ (Å) | 2.44 | 2.42 | 2.42 | 2.45 | 2.87 | 2.47 |

Table II. The segregation energies and magnetic moments of Mn (radius=1.30 Å) at some selected atomic sites near the Σ3 (112) GB in Ge where the dangling bonds are saturated by additional H atoms and the grain boundaries are separated by a vacuum layer.

| | Mn site | 3 | 4 | 6 | 8 |
|---|---|---|---|---|---|
| H-terminated model | $E_{seg}$ (eV) | -0.18 | -0.01 | 0.22 | 0.07 |
| | Magnetic moment ($\mu_B$/Mn) | 3.66 | 3.65 | 3.09 | 3.27 |

TABLE III. The elastic deformation energy $E_{el}$ (eV) induced by Mn at some selected atomic sites near the Σ3 (112) GB in Ge. Also listed are the segregation energies for those sites as presented in Table I. For denotation of Ge sites, see Fig. 3.

| Mn site | 3 | 4 | 6 | 8 |
|---|---|---|---|---|
| $E_{el}$ (eV/Mn) | 0.12 | 0.01 | 0.24 | 0.14 |
| $E_{Seg}$ (eV) | -0.14 | -0.30 | 0.22 | 0.03 |

**Figure Captions**

FIG. 1. (Color online) Step by step construction of a Σ3 (112) GB with rigid body-translation. (a) The symmetrical rigid GB. (b) A translated structure obtained by removing atoms 'A' and 'B', and performing a rigid body translation along <111> and <112>. (c) The translated geometrical model obtained from Fig. 1(b).

FIG. 2. (Color online) Changes in total energy when adding more atomic layers to the Ge (112) slab. All atoms were fixed at the corresponding bulk positions.

FIG. 3. (Color online) Supercell used to model the RBT (2×1) or RBT (1×1) Σ3 (112) GB in Ge. The No. 2 and No. 4 atoms (silver) are three-fold coordinated in the (1×1) model, and will reconstruct in (2×1) model along [110]. The red "I" atom resides at the interstitial site. The atom "B" is viewed as in a bulk environment. The atoms in four outermost Ge layers are fixed at the corresponding bulk positions.

FIG. 4. (Color online) The calculated valence electron density on the (112) plane containing two No. 4 atoms (see Fig. 3), which are near the Σ3 (112) GB in Ge, for both the RBT (1×1) (panel a) and RBT (2×1) (panel b) geometry. Contours start from 0.02 e/a.u.³ and increase successively by a factor of $\sqrt{2}$.

**FIG. 5.** The local magnetic moment of Mn (radius=1.3Å) as a function of the Mn-Ge bond length in MnGe supercell.